\documentclass[12pt]{article}
\usepackage{amsfonts}
\usepackage{amsmath}
\usepackage{amssymb}
\usepackage{amstext}
\usepackage{amsfonts}
\usepackage{graphicx,epsfig}
\usepackage{indentfirst}
\usepackage{hyperref}
\makeatletter
\def\ExtendSymbol#1#2#3#4#5{\ext@arrow 0099{\arrowfill@#1#2#3}{#4}{#5}}
\def\RightExtendSymbol#1#2#3#4#5{\ext@arrow 0359{\arrowfill@#1#2#3}{#4}{#5}}
\def\LeftExtendSymbol#1#2#3#4#5{\ext@arrow 6095{\arrowfill@#1#2#3}{#4}{#5}}
\makeatother

\begin{document}

\title{Construction of Bound Entangled States Based on Permutation Operators
}


\author{Hui Zhao $^{1}$
        Sha Guo $^{1}$
        Naihuan Jing $^{2}$$^{3}$
        Shao-Ming Fei $^{4}$\\
\small \sl $^1$ College of Applied Sciences, Beijing University of Technology, Beijing 100124, China\\
\small \sl $^2$  Department of Mathematics, North Carolina State University, Raleigh, NC 27695, USA\\
\small \sl $^3$ School of Mathematics, South China University of Technology, Guangzhou 510640, China\\
\small \sl $^4$ School of Mathematical Sciences, Capital Normal University, Beijing 100048, China
}

\date{}
\maketitle

\begin{abstract}

We present a construction of new bound entangled states from given bound entangled states
for arbitrary dimensional bipartite systems.
One way to construct bound entangled states is to show that these states are PPT
(positive partial transpose) and violate the range criterion at the same time.
By applying certain operators to given bound entangled states
or to one of the subsystems of the given bound entangled states,
we obtain a set of new states which are both PPT
and violate the range criterion. We show that the derived bound
entangled states are not local unitary equivalent to
the original bound entangled states by detail examples.

\end{abstract}

\section{Introduction}

\label{intro}
Quantum entanglement has played an important role in quantum information processing such as quantum teleportation \cite{Ref1},
quantum cryptography \cite{Ref2}, quantum dense coding \cite{Ref3}, and parallel computing \cite{Ref4}. Among quantum entangled states,
one special type of entangled ones is the bound entangled states. Even though no pure entanglement can be distilled from bound
entangled states they constitute a useful resource in quantum information protocols. They can be helpful for quantum communication
via activation \cite{RefA,RefB}. It was also shown that some bound entangled states can be useful in enhancing teleportation power \cite{RefC},
distilling secure quantum keys \cite{RefD} and reducing communication complexity \cite{RefE}.

As bound entangled states show different characters of entanglement from that of distillable quantum states,
it is of significance to study the structure and learn the characterization of these states.
Considerable efforts have been made to the construction of bound entangled states. Such
constructions provide a deep insight into the structure of entangled states. Meanwhile, many useful tools are introduced in identifying bound entanglement.
The first example of bound entanglement was given by Horodecki \cite{Ref5}. Bound entangled states are also constructed based on unextendible product bases (UPB)
\cite{Ref6} and mutually unbiased bases (MUB) \cite{Ref7,Ref8}. A systematic method to construct high-dimensional bound entangled states was presented in Ref. \cite{Ref9}.
High-dimensional bipartite and multipartite bound entangled states are studied in Ref. \cite{Ref10,Ref11,Ref12,Ref14}. Although many bound entangled
states have been found, the physical character and mathematical structure of bound entangled states are still not well understood.

In this paper, we propose a construction of bound entangled states in any bipartite systems. By using actions
on a given bound entangled states or on one of the subsystems of these states, a series of bound entangled states can be constructed,
which are not local unitary equivalent to the given bound entangled states. The paper is organized as follows:
In Section 2, we demonstrate that the states obtained by applying the operators on the given states are PPT and violate the range criterion.
Examples of bound entangled states are given. In Section 3, we show that the states can be bound entangled by applying
the operators to one of the subsystems of given bound entangled states.
We present some detailed examples of this construction. Conclusions and discussions are given in Section 4.

\section{Construction via operators acting on density matrix}
\label{sec:1}
Let $H$ be an $N$-dimensional complex Hilbert space with an orthonormal basis $|i\rangle$, $i=1,\cdots, N$.
Let $\rho$ be a density matrix defined on $H\otimes H$ with rank($\rho$)=$n\leq N^{2}$. Then $\rho$ can be written as
   \begin{eqnarray*}
     \rho=\sum_{i=1}^{n}\lambda_{i}|\nu_{i}\rangle\langle \nu_{i}|,
   \end{eqnarray*}
 where $|\nu_{i}\rangle$ and $\lambda_{i}$ are the eigenvectors and eigenvalues, respectively. $|\nu_{i}\rangle$ is a normalized bipartite pure state of the form
  \begin{eqnarray*}
      |\nu_{i}\rangle=(a_{1i},\cdots, a_{N^{2}i})^{t},
    \end{eqnarray*}
where $t$ stands for transposition.
Let $P_{mn}$ be the permutation operator that swaps the $m$th and $n$th systems, i.e.
    \begin{eqnarray*}
      P_{mn}|1,2,\cdots,m,\cdots,n,\cdots,N\rangle=|1,2,\cdots,n,\cdots,m,\cdots,N\rangle.
    \end{eqnarray*}
Let $Q_{i}(c)P_{mn}$ be the matrix obtained by multiplying the $i$th row or column of $P_{mn}$ by $c$, where $c\neq 0,1$, $c\in\mathcal{R}$ is a real number.

\textbf{Theorem 1.} Suppose that
  the state $\rho$ is PPT and not satisfying the range criterion. Then
   for any $Q=Q_{i}(c)P_{mn}$, 
   $c\neq 0,1\in \mathcal{R}$ and $i\neq m, \ n$, the state $\rho'=(I\otimes Q)\rho(I\otimes Q)^{\dagger}$ is also bound entangled.

\textbf{Proof:} Since $(I\otimes Q)^{T_{2}}=I\otimes Q$, $(I\otimes Q)^{\dagger}=(I\otimes Q)^{t}$ and the congruent transformation dose not change
 positive semi-definiteness of the matrix, $\rho'^{T_{2}}$ is positive semi-definite, where $T_{2}$ denotes the transposition with respect to the
 second system. Hence $\rho'$ is PPT. Any vector $\mu$ in the range
of $\rho$ is a vector of the form
  \begin{eqnarray*}
     \mu = (B_{11}, \cdots,B_{1N},\cdots,B_{N1},\cdots,B_{NN} )^{t},
  \end{eqnarray*}
where $B_{ij}\in \mathcal{C}.$  As $\rho'=(I\otimes Q)\rho(I\otimes Q)^{\dagger}$, we see
  that the corresponding vector $\mu'$ in the range of $\rho'$ is given by
$$\mu'=(B_{1'1'},\cdots,cB_{1'i'},\cdots, B_{1'N'},\cdots, B_{N'1'},\cdots, B_{N'N'})^t,$$
where $P_{mn}(i)=i'$.

Since the state $\rho$ violates the range criterion, there exist a basis
$\{\mu_{1},\cdots,\mu_{q}\}$ of range$(\rho)$ such that their partial complex conjugations with
respect to the second system do not span range$(\rho^{T_{2}})$. That is, there is a vector $\mu_{0}$ belonging to
the range of $\rho$ which is linearly independent from the vectors spanning the range of $\rho^{T_{2}}$.
Since $I\otimes Q$ is reversible,
the vectors $(I\otimes Q)\mu_{1},\cdots,(I\otimes Q)\mu_{q}$ span
the range of $\rho'$, and the vector $(I\otimes Q)\mu_{0}$ belonging to the range of $\rho'^{T_{2}}$ is also linearly independent
from the partial complex conjugated vectors. Hence the state $\rho'$ is bound entangled.

\textbf{Remark 1.} According to Ref. [20], two density matrices are equivalent under local unitary transformations if there exists an
ordering of the corresponding eigenstates such that the following invariants have the same values for
both density matrices:
$$J^{s}(\rho)=Tr(\rho^{s}), \quad \widetilde{\Omega}(\rho),\quad \widetilde{\theta}(\rho), \quad\widetilde{X}(\rho), \quad \widetilde{Y}(\rho),$$
where $i, j, k, s=1,\cdots,N^{2}$, with entries give by $\widetilde{\Omega}(\rho)_{ij}=Tr(\rho_i\rho_j)$,
$\widetilde{\theta}(\rho)_{ij}=Tr(\theta_i\theta_j)$, $\widetilde{X}(\rho)_{ijk}=Tr(\rho_i\rho_j\rho_k)$,
$\widetilde{Y}(\rho)_{ijk}=Tr(\theta_i\theta_j\theta_k)$, where $\rho_i=Tr_2|v_i\rangle\langle v_i|$, $\theta_i=Tr_1|v_i\rangle\langle v_i|$.
Therefore if any of the above invariants are not equal for two density matrices,
then they are not equivalent
under local unitary transformations. By choosing appropriate operators $Q$
one sees that the bound entangled states constructed in Theorem 1 are not
local unitary equivalent to the original bound entangled states.

\textbf{Remark 2.} If $\dim{H_{1}}\neq \dim{H_{2}}$,
by using the same approach
one can also get similar results for bipartite states in $H_1\otimes H_2$.

\textbf{Remark 3.} Instead of the operator $Q$ in the Theorem 1, if one uses
$Q=Q_{k}(c)\prod\limits_{ij} P_{ij}$, $k\neq i\neq j$, where $\prod\limits_{ij}$ denotes the
product of a finite number matrices $P_{ij}$, similar results can be obtained.

Next we will give some examples.

\textbf{\textbf{Example 1.}} The spectral decomposition of the state $\rho$ \cite{Ref14} in $3\otimes3$ systems can be expressed as,
\begin{equation} \label{1}
 \rho=\frac{(1-\varepsilon)}{3}(|f_{1}\rangle\langle f_{1}|+|f_{2}\rangle\langle f_{2}|+|f_{3}\rangle\langle f_{3}|)+\frac{\varepsilon}{2}(|f_{4}\rangle\langle f_{4}|+|f_{5}\rangle\langle f_{5}|),
\end{equation} where $0<\varepsilon\leq\frac{2}{5}$,
\begin{eqnarray*}
    |f_{1}\rangle &=& (1,0,0,0,0,0,0,0,0)^{t}, \\
    |f_{2}\rangle &=& (0,0,0,0,1,0,0,0,0)^{t}, \\
    |f_{3}\rangle &=& (0,0,0,0,0,0,0,0,1)^{t}, \\
    |f_{4}\rangle &=& (0,\frac{1}{\sqrt{2}},0,-\frac{1}{\sqrt{2}},0,0,0,0,0)^{t}, \\
    |f_{5}\rangle &=& (0,0,\frac{1}{\sqrt{2}},0,0,0,-\frac{1}{\sqrt{2}},0,0)^{t}.
  \end{eqnarray*}
Let $Q=Q_3(c)P_{12}$, then $\rho'=(I\otimes Q)\rho(I\otimes Q)^{\dagger}$ is bound entangled.

According to \cite{Ref14}, the state $\rho$ is PPT and violates the range criterion. From Theorem 1 $\rho'$ is PPT too.
Any vector of range$(\rho)$ can be represented as
\begin{eqnarray*}
     \mu = (A,B,C,-B,D,0,-C,0,E )^{t},
  \end{eqnarray*}where $A,B,C,D,E \in \mathcal{C}.$ Any vector of range$(\rho')$ can be expressed as
\begin{eqnarray*}
     \mu' = (B,A, C,D,-B,0,0,-C,E )^{t}.
  \end{eqnarray*}
According to \cite{Ref14}, vectors $\mu_{1}=(a_{1},0,0)^{t}\otimes(b_{1},0,0)^{t},\ \ \mu_{2}=(0,a_{2},0)^{t}\otimes(0,b_{2},0)^{t},\ \  \mu_{3}=(0,0,a_{3})^{t}\otimes(0,0,b_{3})^{t},\ \ \mu_{4}=(a_{1},0,a_{3})^{t}\otimes(b_{1},0,b_{3})^{t},\ \ \mu_{5}=(a_{1},a_{2},0)^{t}\otimes(b_{1},b_{2},0)^{t}$ span the range of $\rho$, while their
partial complex conjugations  on the second system do not span the range of $\rho^{T_{2}}$, as the vector $\mu_{0}=(1,0,0)\otimes(0,1,0)^{t}$ of range$(\rho^{T_{2}})$
is linearly independent from any basis vectors of range$(\rho)$.
For $\rho'$ we can also get that the partial complex conjugations with respect to the second system of
the following vectors do not span the range of $\rho'^{T_{2}}$, since the vector $(I\otimes Q)\mu_{0} = (1,0,0)^{t}\otimes(1,0,0)^{t}$ of
range$(\rho'^{T_{2}})$ is linearly independent from these vectors,
\begin{eqnarray*}
    (I\otimes Q)\mu_{1} &=& (a_{1},0,0)^{t}\otimes(0,b_{1},0)^{t},  \\
   (I\otimes Q)\mu_{2} &=& (0,a_{2},0)^{t}\otimes(b_{2},0,0)^{t},  \\
   (I\otimes Q)\mu_{3} &=& (0,0,a_{3})^{t}\otimes(0,0,cb_{3})^{t},  \\
   (I\otimes Q)\mu_{4} &=& (a_{1},0,a_{3})^{t}\otimes(0,b_{1},cb_{3})^{t},  \\
    (I\otimes Q)\mu_{5} &=& (a_{1},a_{2},0)^{t}\otimes(b_{2},b_{1},0)^{t}.
    \end{eqnarray*}
Hence $\rho'$ is bound entangled either. Since the invariants $\theta(\rho)_{3,3}=1,\ \theta(\rho')_{3,3}=c^{4}$,
i.e. $\theta(\rho)\neq\theta(\rho')$, $\rho$ and $\rho'$ are not local unitary equivalent.

\textbf{\textbf{Example 2.}} The spectral decomposition of the state $\rho$ in $2\otimes8$ systems \cite{Ref16} has the form,
\begin{equation}\label{2}
\rho=\frac{(1-\varepsilon)}{4}\sum_{i=1}^{4}|\eta_{i}\rangle\langle\eta_{i}|+\frac{\varepsilon}{4}\sum_{i=5}^{8}|\eta_{i}\rangle\langle\eta_{i}|,\ \ \ 0<\varepsilon\leq\frac{1}{2},
\end{equation}
where\begin{eqnarray*}
    |\eta_{1}\rangle &=& (1,0,0,0,0,0,0,0,0,0,0,0,0,0,0,0)^{t}, \\
    |\eta_{2}\rangle &=& (0,0,0,0,0,1,0,0,0,0,0,0,0,0,0,0)^{t}, \\
    |\eta_{3}\rangle &=&  (0,0,0,0,0,0,0,0,0,0,1,0,0,0,0,0)^{t},\\
    |\eta_{4}\rangle &=& (0,0,0,0,0,0,0,0,0,0,0,0,0,0,0,1)^{t}, \\
    |\eta_{5}\rangle &=& (0,\frac{1}{\sqrt{2}},0,0,-\frac{1}{\sqrt{2}},0,0,0,0,0,0,0,0,0,0,0)^{t}, \\
    |\eta_{6}\rangle &=& (0,0,\frac{1}{\sqrt{2}},0,0,0,0,0,-\frac{1}{\sqrt{2}},0,0,0,0,0,0,0)^{t},\\
     |\eta_{7}\rangle &=&(0,0,0,0,0,0,0,\frac{1}{\sqrt{2}},0,0,0,0,0,-\frac{1}{\sqrt{2}},0,0)^{t}, \\
    |\eta_{8}\rangle &=& (0,0,0,0,0,0,0,0,0,0,0,\frac{1}{\sqrt{2}},0,0,-\frac{1}{\sqrt{2}},0)^{t}.
  \end{eqnarray*}
Firstly we prove that the state $\rho$ is PPT and violates the range criterion. $\rho^{T_{2}}$ is a nonzero Hermitian row diagonally matrix
when $ 0<\varepsilon\leq\frac{1}{2}$, thus $\rho^{T_{2}}$ is positive semidefinite. Any vector of range$(\rho)$ can be written as
\begin{eqnarray*}
     \mu = (A,B,C,0,-B,D,0,E,-C,0,F,G,0,-E,-G,H )^{t},
  \end{eqnarray*} where $A,B,C,D,E,F,G,H \in \mathcal{C}.$ If $\rho$ is separable,
\begin{eqnarray*}
     \mu_{sep} = (b_{1},b_{2} )^{t}\otimes (c_{1},\cdots,c_{8}) ^{t}.
  \end{eqnarray*}
Thus we see that the following vectors span range$(\rho)$
  \begin{eqnarray*}
   \mu_{1} &=& (b_{1},b_{2})^{t}\otimes(c_{1},0,c_{3},0,0,c_{6},0,c_{8})^{t},  \\
    \mu_{2} &=& (b_{1},0)^{t}\otimes(c_{1},c_{2},0,0,c_{5},c_{6},0,0)^{t},  \\
    \mu_{3} &=& (0,b_{2})^{t}\otimes(0,0,c_{3},c_{4},0,0,c_{7},c_{8})^{t}.
  \end{eqnarray*}
Since the vector $\mu_{0}=(1,0)^{t}\otimes(0,0,1,0,0,0,0,0)^{t}\in \rho^{T_{2}}$ is linearly independent from
the vectors $\mu_{1}^{*2}, \mu_{2}^{*2}, \mu_{3}^{*2}$, $\rho$ is entangled. Therefore, $\rho$ is bound entangled.

Let $Q=Q_{3}(c)P_{12}P_{78}$. Then $\rho'=(I\otimes Q)\rho(I\otimes Q)^{\dagger}$. It follows from Theorem 1 that $\rho'$ is PPT. Any vector of range$(\rho')$ is of the following form
\begin{eqnarray*}
     \mu = (B,A,cC,0,-B,D,E,0,0,-C,F,G,0,-E,H,-G )^{t}.
  \end{eqnarray*}
We get that
  \begin{eqnarray*}
    (I\otimes Q)\mu_{1} &=& (b_{1},b_{2})^{t}\otimes(0,c_{1},cc_{3},0,0,c_{6},c_{8},0)^{t}, \\
   (I\otimes Q)\mu_{2} &=& (b_{1},0)^{t}\otimes(c_{2},c_{1},0,0,c_{5},c_{6},0,0)^{t},  \\
   (I\otimes Q)\mu_{3} &=& (0,b_{2})^{t}\otimes(0,0,cc_{3},c_{4},0,0,c_{8},c_{7})^{t},  \\
   (I\otimes Q)\mu_{0} &=&(1,0)^{t}\otimes(0,0,c,0,0,0,0,0)^{t}.
  \end{eqnarray*}
So $\rho'$ is also bound entangled. Since $\theta(\rho)_{6,7}=0,\ \theta(\rho')_{6,7}=\frac{1}{4}+\frac{1}{4}c^{2}$, then $\theta(\rho)\neq \theta(\rho')$,
and $\rho$ and $\rho'$ are not local unitary equivalent.

In \cite{Ref14}, we have presented a class of bound entangled states in $3k\otimes3k$ quantum systems. Using Theorem 1, we can construct new bound entangled states
from these $3k\otimes3k$ bound entangled states.

\textbf{\textbf{Example 3.}} The spectral decomposition of the bound entangled state $\rho$ \cite{Ref14} in $3k\otimes3k$ quantum systems is written as follows:
\begin{eqnarray*}
\rho=\frac{\varepsilon}{2}\sum_{i=1}^{2}|\chi_{i}\rangle\langle\chi_{i}|+\frac{(1-\varepsilon)}{7k^{2}-4k}\sum_{i=3}^{7k^{2}-4k+2}|\chi_{i}\rangle\langle\chi_{i}|,\ \ \ 0<\varepsilon\leq\frac{2}{7k-2},
\end{eqnarray*}
where $|\chi_{1}\rangle=|\phi_{1}\rangle$ and $|\chi_{2}\rangle=|\phi_{2}\rangle$ are the linearly independent eigenvectors corresponding to the eigenvalue $\frac{\varepsilon}{2}$, with $|\phi_{1}\rangle=(0,b,0,0,\cdots,0,-b,0,0,0,\cdots,0,\\0,b,0,0,\cdots,0,-b,0,0,0,\cdots,0)^t,$
\ $|\phi_{2}\rangle=(0,0,b,0,\cdots,0,0,\cdots,0,-b,0,0,0,\cdots,\\0,0,0,b,0\cdots,0,0,\cdots,0,-b,0,0)^t,\ |b|^{2}=\frac{1}{2k}$.
$|\chi_{i}\rangle,\ i=3,\cdots,7k^{2}-4k+2$, are the linearly independent
eigenvectors of $L_{3k}$ with eigenvalue of $\frac{1-\varepsilon}{7k^{2}-4k}$, where $L_{3k}$ is a $9k^{2}\times9k^{2}$ matrix having the following nonzero entries:

$(L_{3k})_{(m-1)\times3k+m,(m-1)\times3k+m}=\frac{1}{7k^{2}-4k},\ \ \ \  m=1,2,\cdots,3k.$

$(L_{3k})_{(3m-l)\times3k+3n-l',(3m-l)\times3k+3n-l'}=\frac{1}{7k^{2}-4k},\ \ \ \ m,n=1,2,\cdots,k,\  m \neq n$.\ \    $l'=\left\{\begin{array}{ccc}
                               {0,2} & , & l=1. \\
                               {1,2} & , & l=2.\\
                               {0,1,2} & , & l=3.\\
                             \end{array}
\right.$

Take $Q=P_{3(c)}P_{12} $. We claim that the state $\rho'=(I\otimes Q)\rho(I\otimes Q)^{\dagger}$ is bound entangled.
In fact, since $\rho$ is PPT, $\rho'$ is also PPT. By \cite{Ref13}, the following vectors form a basis of the range of $\rho$:

$|\psi_{3m-2,3m-2}\rangle =(0,\cdots,0,a_{3m-2},0,\cdots,0)^t\otimes(0,\cdots,0,b_{3m-2},0,\cdots,0)^t,$

$|\psi_{3m-2,3n-2}\rangle =(0,\cdots,0,a_{3m-2},0,\cdots,0)^t\otimes(0,\cdots,0,b_{3n-2},0,\cdots,0)^t,$

$|\psi_{3m-2,3n-1}\rangle = (0,\cdots,0,a_{3m-2},0,\cdots,0)^t\otimes(0,\cdots,0,b_{3n-1},0,\cdots,0)^t,$

$|\psi_{3m-2,3n}\rangle=(0,\cdots,0,a_{3m-2},0,\cdots,0)^t\otimes(0,\cdots,0,b_{3n},0,\cdots,0)^t,$

$|\psi_{3m-1,3m-1}\rangle =(0,\cdots,0,a_{3m-1},0,\cdots,0)^t\otimes(0,\cdots,0,b_{3m-1},0,\cdots,0)^t,$

$|\psi_{3m-1,3n-2}\rangle = (0,\cdots,0,a_{3m-1},0,\cdots,0)^t\otimes(0,\cdots,0,b_{3n-2},0,\cdots,0)^t,$

$|\psi_{3m-1,3n-1}\rangle = (0,\cdots,0,a_{3m-1},0,\cdots,0)^t\otimes(0,\cdots,0,b_{3n-1},0,\cdots,0)^t, $

$|\psi_{3m,3m}\rangle= (0,\cdots,0,a_{3m},0,\cdots,0)^t\otimes(0,\cdots,0,b_{3m},0,\cdots,0)^t, $

$|\psi_{3m,3n-2}\rangle = (0,\cdots,0,a_{3m},0,\cdots,0)^t\otimes(0,\cdots,0,b_{3n-2},0,\cdots,0)^t,$

$|\psi_{3m,3n}\rangle = (0,\cdots,0,a_{3m},0,\cdots,0)^t\otimes(0,\cdots,0,b_{3n},0,\cdots,0)^t,$

$|\psi_{k}\rangle = (a_{1},0,a_{3},\cdots,a_{3k-2},0,a_{3k})^{t}\otimes(b_{1},0,b_{3},\cdots,b_{3k-2},0,b_{3k})^{t},$

$|\psi_{kk}\rangle =(a_{1},a_{2},0,\cdots,a_{3k-2},a_{3k-1},0)^{t}\otimes(b_{1},b_{2},0,\cdots,b_{3k-2},b_{3k-1},0)^{t},$

\noindent where $m,n=1,\cdots,k,\ n\neq m$, $a_{1},\cdots,a_{3k},b_{1},\cdots,b_{3k}\in \mathcal{C}.$
The vector $|\psi_{0}\rangle=(1,0,\cdots,0)^t\otimes(0,1,0,\cdots,0)^t$ of range$(\rho^{T_{2}})$ is
also linearly independent from the vectors obtained by taking partial complex conjugation on the above vectors.

According to Theorem 1 we know that the vectors
$(I\otimes Q)|\psi_{3m-2,3m-2}\rangle$,\ $(I\otimes Q)|\psi_{3m-2,3n-2}\rangle$,\ $(I\otimes Q)|\psi_{3m-2,3n-1}\rangle$,\ $(I\otimes Q)|\psi_{3m-2,3n}\rangle$,\ $(I\otimes Q)|\psi_{3m-1,3m-1}\rangle$,\ $(I\otimes Q)|\psi_{3m-1,3n-2}\rangle$,\ $(I\otimes Q)|\psi_{3m-1,3n-1}\rangle$,\ $(I\otimes Q)|\psi_{3m,3m}\rangle$,\ $(I\otimes Q)|\psi_{3m,3n-2}\rangle$,\ $(I\otimes Q)|\psi_{3m,3n}\rangle$,\ $(I\otimes Q)|\psi_{k}\rangle$,\ $(I\otimes Q)|\psi_{kk}\rangle$ span
the range of $\rho'$. Performing the partial complex conjugations with respect to the second system, we get that the resulting vectors do not span the range of $\rho'^{T_{2}}$, since the range vector $(I\otimes Q) |\psi_{0}\rangle$ of $\rho'^{T_{2}}$ is linearly independent from the resulting vectors. Hence $\rho'$ is bound entangled. Moreover, since $\theta(\rho)_{2,2}=\frac{1}{2k}$, $\theta(\rho')_{2,2}=\frac{1}{2k}+(c^{4}-1)\frac{1}{4k^{2}}$, hence $\theta(\rho)\neq \theta(\rho')$. Therefore, $\rho$ and $\rho'$ are not local unitary equivalent.

\section{Construction by action on bases of the density matrices}
\label{sec:2}
In this section, we consider construction of bound entangled states
based on changing bases of the density matrices. We restrict ourselves to permutation operators
invariant under $T_2$ and set $P^{(1)}=\{P_{mn}\in P \mid P_{mn}^{T_{2}}=P_{mn}\}$.

Let $\sigma=\sum_{i=1}^{n}\lambda_{i}|\nu_{i}\rangle\langle \nu_{i}|$ be a density matrix under spectral decomposition.
Suppose there is a permutation operator $P_{mn}\in P^{(1)}$ leaving all
eigenvectors $|\nu_{j}\rangle$ invariant except for possibly $|\nu_{i}\rangle$. That is, the
components $a_{mj}$ of $|\nu_{j}\rangle$
satisfy $ a_{mj}= a_{nj}$ for all $j=1,\cdots, i-1,i+1,\cdots, n.$
Then we have the following theorem.

\textbf{Theorem 2.}
If the density matrix $\sigma$ constructed as above is PPT and dose not satisfy the range criterion, then
\begin{eqnarray*}
     \sigma'=\sum_{j\neq i}\lambda_{j}|\nu_{j}\rangle\langle \nu_{j}|+\lambda_{i}P_{mn}|\nu_{i}\rangle\langle \nu_{i}|P_{mn}^{\dagger}
   \end{eqnarray*}
is bound entangled.

\textbf{Proof:}
By assumption, $\sigma'=P_{mn}\sigma P_{mn}^{\dagger}=P_{mn}\sigma P_{mn}$.
Since $P_{mn}^{T_{2}}=P_{mn}$ and $\sigma^{T_{2}}$ is positive semi-definite, $\sigma'^{T_{2}}=P_{mn}\sigma^{T_{2}}P_{mn}$ is positive semi-definite. Hence $\sigma'$ is PPT.

Any vector $\mu$ of range$(\sigma)$ can be written as
\begin{eqnarray*}
 \mu = (A_{11},\cdots,A_{1N},\cdots,A_{N1},\cdots,A_{NN})^{t}.
\end{eqnarray*}
Under the action of $P_{mn}$,
the corresponding vector $\mu'$ in range$(\sigma')$ becomes
$$\mu'=(A_{1'1'},\cdots, A_{1'N'},\cdots, A_{N'1'},\cdots, A_{N'N'})^t, $$
where $P_{mn}(i)=i'$. Suppose that the vectors $\mu_{1},\mu_{2},\cdots,\mu_{q}$ span the range of $\sigma$, but their partial complex
conjugations with respect to the second system do not span the range of $\sigma^{T_{2}}$. That is, there is a vector $\mu_{0}$ of range$(\sigma^{T_{2}})$ which
is linearly independent from these conjugated vectors.
Since $P_{mn}$ is reversible, there are vectors
$P_{mn}\mu_{1},\cdots,P_{mn}\mu_{l},$ $l\leq q,$ span the range of $\sigma'$, and the vector
$P_{mn}\mu_{0}$ of range$(\sigma'^{T_{2}})$ is also linearly independent from these spanning vectors under
partial complex conjugation on the second system. Thus the state $\sigma'$ is bound entangled.

\textbf{Remark 4.} If $\dim{H_{1}}\neq dim{H_{2}}$, the similar result still holds for states in $H_1\otimes H_2$.

\textbf{Remark 5.} According to \cite{Ref15}, by using the local unitary invariants
$\widetilde{\Omega}(\sigma)$, $\widetilde{\theta}(\sigma)$, $\widetilde{X}(\sigma)$, $\widetilde{Y}(\sigma)$,
together with the condition $J^{s}(\sigma)=Tr(\sigma^{s}), s=1,\cdots,N^{2}$, one can verify that, by choosing appropriate operators,
the derived states are not local unitary equivalent to the original states.

\textbf{Example 4:} Consider the state Eq.(\ref{1}) in Example 1. Let $P_{46}$ act on the eigenvector $ |f_{4}\rangle$. We have
 \begin{eqnarray*}
   \sigma' = P_{46} \rho P_{46}^{\dagger}=\frac{(1-\varepsilon)}{3}(|f_{1}\rangle\langle f_{1}|+|f_{2}\rangle\langle f_{2}|+|f_{3}\rangle\langle f_{3}|)+&&\nonumber\\ \frac{\varepsilon}{2}(P_{46}|f_{4}\rangle\langle f_{4}|P_{46}^{\dagger}+|f_{5}\rangle\langle f_{5}|), \ 0<\varepsilon\leq\frac{2}{5}.
 \end{eqnarray*}
Obviously $\sigma'$ is PPT. Vectors $\mu_{1}=(a_{1},0,0)^{t}\otimes(b_{1},0,0)^{t},\ \mu_{2}=(0,a_{2},0)^{t}\otimes(0,b_{2},0)^{t},\ \mu_{3}=(0,0,a_{3})^{t}\otimes(0,0,b_{3})^{t},\ \mu_{4}=(a_{1},0,a_{3})^{t}\otimes(b_{1},0,b_{3})^{t},\ \mu_{5}=(a_{1},a_{2},0)^{t}\otimes(b_{1},b_{2},0)^{t}$ span the range of $\rho$. However, the partial complex conjugations of these vectors do not span the range of $\rho^{T_{2}}$, as the vector $\mu_{0}=(1,0,0)\otimes(0,1,0)^{t}$ is linearly independent from these vectors. According to Theorem 2, we can get
 \begin{eqnarray*}
    P_{46}\mu_{1} &=& (a_{1},0,0)^{t}\otimes(b_{1},0,0)^{t},  \\
    P_{46}\mu_{2} &=& (0,a_{2},0)^{t}\otimes(0,b_{2},0)^{t},  \\
    P_{46}\mu_{3} &=& (0,0,a_{3})^{t}\otimes(0,0,b_{3})^{t},  \\
    P_{46}\mu_{4} &=& (a_{1},0,a_{3})^{t}\otimes(b_{1},0,b_{3})^{t}.
    \end{eqnarray*}

The vector $P_{46}\mu_{0} = (1,0,0)^{t}\otimes(0,1,0)^{t}$ belonging to the range of $\sigma'^{T_{2}}$ is linearly independent from
the above vectors under partial complex conjugation on the second system. Hence $\sigma'$ is also bound entangled. Since
$\theta(\rho)_{3,4}=0,\  \theta(\sigma')_{3,4}=\frac{1}{2}$, i.e. $\theta(\rho)\neq\theta(\sigma'),$  $\rho$ and $\sigma'$ are not local unitary equivalent.

\textbf{Example 5:} Consider the state Eq.(\ref{2}) in Example 2.
Let $P_{24}$ act on the eigenvector $|\eta_{5}\rangle$. Then the new state
\begin{eqnarray*}
  \sigma' =P_{24} \rho P_{24}^{\dagger} =\frac{(1-\varepsilon)}{4}\sum_{i=1}^{4}|\eta_{i}\rangle\langle\eta_{i}|+\frac{\varepsilon}{4}(P_{24}|\eta_{5}\rangle\langle\eta_{5}|P_{24}^{\dagger}+\sum_{i=6}^{8}|\eta_{i}\rangle\langle\eta_{i}|),\ 0<\varepsilon\leq\frac{1}{2},
\end{eqnarray*} is bound entangled. Namely, $\sigma'$ is also PPT. Moreover,
\begin{eqnarray*}
    P_{24}\mu_{1} &=& (b_{1},b_{2})^{t}\otimes(c_{1},0,c_{3},0,0,c_{6},0,c_{8})^{t},  \\
    P_{24}\mu_{2} &=& (b_{1},0)^{t}\otimes(c_{1},0,0,c_{2},c_{5},c_{6},0,0)^{t}, \\
    P_{24}\mu_{3} &=& (0,b_{2})^{t}\otimes(0,0,c_{3},c_{4},0,0,c_{7},c_{8})^{t}.
  \end{eqnarray*}
The vector $P_{24}\mu_{0} = (1,0)^{t}\otimes(0,0,1,0,0,0,0,0)^{t}$ of range$(\sigma'^{T_{2}})$ is linearly independent from
the above vectors under partial complex conjugation on the second system.  Therefore $\sigma'$ is also bound entangled.
Since $\theta(\rho)_{4,5}=0,\ \theta(\sigma')_{4,5}=\frac{1}{2}$, then $\theta(\rho)\neq \theta(\sigma')$. Thus $\rho$ and $\sigma'$ are not local unitary equivalent.

\textbf{Example 6:} Consider the bound entangled states $\rho$ defined in Example 3.
Let operator $P_{(3k+1)(3k+3)}$ act on the eigenvector $|\chi_{1}\rangle$. Then the following state is also bound entangled,
 \begin{eqnarray*}
\sigma'=\frac{\varepsilon}{2}(P_{(3k+1)(3k+3)}|\chi_{1}\rangle\langle\chi_{1}|P_{(3k+1)(3k+3)}^{\dagger}+|\chi_{2}\rangle\langle\chi_{2}|) &&\nonumber\\ +\frac{(1-\varepsilon)}{7k^{2}-4k}\sum_{i=3}^{7k^{2}-4k+2}|\chi_{i}\rangle\langle\chi_{i}|,\ \ \ 0<\varepsilon\leq\frac{2}{7k-2}.
\end{eqnarray*}
This can be seen as follows.
Since $\sigma'=P_{(3k+1)(3k+3)}\rho P_{(3k+1)(3k+3)}^{\dagger}$, $\sigma'$ is PPT. According to Theorem 2, we have that the vectors
$P_{(3k+1)(3k+3)}|\psi_{3m-2,3m-2}\rangle$,\ $P_{(3k+1)(3k+3)}|\psi_{3m-2,3n-2}\rangle$,\ $P_{(3k+1)(3k+3)}|\psi_{3m-2,3n-1}\rangle$,\
$ P_{(3k+1)(3k+3)}|\psi_{3m-2,3n}\rangle$,\ $ P_{(3k+1)(3k+3)}|\psi_{3m-1,3m-1}\rangle$,\  $ P_{(3k+1)(3k+3)}|\psi_{3m-1,3n-1}\rangle$,\
$ P_{(3k+1)(3k+3)}|\psi_{3m,3m}\rangle$,\ \ $P_{(3k+1)(3k+3)}|\psi_{3m-1,3n-2}\rangle$,\ \ $P_{(3k+1)(3k+3)}|\psi_{3m,3n-2}\rangle$,\
$ P_{(3k+1)(3k+3)}|\psi_{3m,3n}\rangle$,\\ $ P_{(3k+1)(3k+3)}|\psi_{k}\rangle$ span the range of $\sigma'$.  However, the vector
$P_{(3k+1)(3k+3)}|\psi_{0}\rangle$ of range$(\sigma'^{T_{2}})$ is still linearly independent from
the above vectors under partial complex conjugation on the second system.  Thus $\sigma'$ is bound entangled.
Since $\theta(\rho)_{1,3}=\frac{1}{2k}$, $\theta(\sigma')_{1,3}=0$, i.e. $\theta(\rho)\neq \theta(\sigma')$,
$\rho$ and $\sigma'$ are not local unitary equivalent.

\section{Conclusion and Discussion}
\label{sec:3}
We have presented a new construction of bound entangled states from given bound entangled states.
The key operation is based on suitable action on the given bound entangled states.
We have also generalized the method to allow action on the subsystems of the given states.  The
approach gives rise to a series of bound entangled states from a given entangled one.
Moreover, by choosing appropriate operators, the derived bound entangled states are shown to be
local unitary inequivalent to the original bound entangled states.


\section*{ACKNOWLEDGMENTS}
This work is supported by the China Scholarship Council, Simons Foundation 198129, the National Natural Science Foundation of China (11101017, 11275131, 11281137, 11271138 and 11531004), Beijing Natural Science Foundation Program and Scientific Research Key
Program of Beijing Municipal Commission of Education (KZ201210028032) and the Importation
and Development of High-Caliber Talent Project of Beijing Municipal Institutions
(CITTCD201404067).



\end{document}